\documentstyle[12pt,equations,epsf]{article}

\begin{document}

\def\beq{\begin{equation}}
\def\eeq{\end{equation}}
\def\bea{\begin{eqnarray}}
\def\eea{\end{eqnarray}}
\def\bit{\begin{itemize}}
\def\eit{\end{itemize}}
\def\wtil{\widetilde}
\def\what{\widehat}

\def\vev#1{\langle #1 \rangle}

\def\mev{~\mbox{MeV}}
\def\gev{~\mbox{GeV}}
\def\kev{~\mbox{keV}}
\def\tev{~\mbox{TeV}}
\def\eps{\epsilon}
\def\mw{m_W}
\def\mz{m_Z}
\def\leff{{\cal L}_{\rm eff}}
\def\mplanck{M_{\rm Pl}}
\def\nn{\nonumber}
\def\cntwo{\wt\chi_2^0}
\def\wt{\widetilde}
\def\gl{\wt g}
\def\mgl{m_{\gl}}
\def\lsim{\mathrel{\raise.3ex\hbox{$<$\kern-.75em\lower1ex\hbox{$\sim$}}}}
\def\gsim{\mathrel{\raise.3ex\hbox{$>$\kern-.75em\lower1ex\hbox{$\sim$}}}}
\def\fbi{~{\rm fb}^{-1}}
\def\etmiss{/ \hskip-6pt E_T \hskip6pt}
\def\cmone{\tilde{\chi}_1^-}
\def\mcpmone{m_{\cpmone}}
\def\cpone{\tilde{\chi}_1^+}
\def\cpmone{\tilde{\chi}_1^\pm}
\def\cnone{\tilde{\chi}_1^0}
\def\mcnone{m_{\cnone}}
\def\gev{~{\rm GeV}}
\def\mev{~{\rm MeV}}
\def\dmchi{\Delta m_{\tilde{\chi}_1}}
\def\epem{e^+e^-}
\def\anti{\overline}
\def\lam{\lambda}
\def\gam{\gamma}
\def\del{\delta}
\def\dbar{\overline D}
\def\dhat{\widehat D}
\def\htil{\widetilde h}
\def\phitil{\widetilde \phi}
\def\call{{\cal L}}
\def\lmix{\call_{\rm mix}}
\def\lkin{\call_{\rm kin}}
\def\lphi{{\call_\Phi}}
\def\half{{1\over 2}}
\def\quarter{{1\over 4}}
\def\vtot{V_{\rm tot}}
\def\vtotbar{\overline V_{\rm tot}}
\def\veff{V_{\rm eff}}
\def\vnorm{V_{\rm S}}
\def\sphys{s_{\rm phys}}
\def\msphys{m_{\sphys}}
\def\Phip{\Phi^\prime}
\def\fy{f_{\rm Y}}
\def\ldel{L}

\font\fortssbx=cmssbx10 scaled \magstep2
\hbox to \hsize{
$\vcenter{
\hbox{\fortssbx University of California - Davis}\medskip
%\hbox{\fortssbx University of Wisconsin - Madison}
}$
\vspace*{1.2cm}
$\vcenter{
\hbox{\bf UCD-99-18} 
\hbox{\bf IFT-23-99}
\hbox{\bf hep-ph/9910456}
\hbox{October, 1999}
}$
}
%}

\begin{center}
{\large\bf
Kaluza-Klein Excitations and Electroweak Symmetry Breaking
\\}
\rm
\vskip2pc
{\bf B. Grzadkowski$^{a,b}$ and J.F.Gunion$^a$}\\
\medskip
\small\it
$^a$ Davis Institute for High Energy Physics, 
University of California at Davis,\\
Davis, CA 95616, USA\\
$^b$ Inst. for Theoretical Physics, Warsaw University, Warsaw, Poland
\\
\end{center}
\vskip .5cm
\begin{abstract}
We explore the possibility that the Kaluza-Klein graviton states
induce electroweak symmetry breaking. We also demonstrate
that electroweak
symmetry breaking could have a large impact on KK phenomenology.
\end{abstract}

\bigskip
Recently, it has become clear that quantum gravity as described using
extra dimensions, and the associated Kaluza-Klein
excitations, could have effects at scales far below the measured
Planck mass~\cite{original}. 
In the most popular approach~\cite{arkanietal}, ordinary
particles are confined on a brane (having three spatial
dimensions) with gravity propagating in the bulk,
a situation that can be realized in several string models~\cite{gbulk}.
There has been an outpouring of phenomenological work based
on these ideas; see \cite{hanetal,wellsetal,hewett} and references
therein.
Here, we point out that the KK modes could have a dramatic impact on 
electroweak symmetry breaking.  Indeed, they could provide the EWSB
mechanism. In turn, EWSB could have a large impact on KK phenomenology.

\section{Electroweak Symmetry Breaking}
\medskip

We consider a theory with $\del$ extra dimensions, for which we begin with the
action
\beq
S=-{1\over 2\what\kappa^2}\int d^4x\, d^\del y\sqrt{-\what g}\,R
+{1\over V_\del}\int d^4x\, d^\del y \sqrt{-\what g}\,\Lambda
+\sum_{\rm fields}\int d^4x  \sqrt{-\what g_4}\,\call_{\rm fields}(x)\,,
\label{action}
\eeq
where $\what\kappa$ is the reduced Planck mass in $4+\del$ dimensions,
$V_\del=\ldel ^\del$ is the volume associated with the extra compactified
dimensions, $\kappa=V_\del^{-1/2}\what\kappa$ is the
usual reduced Planck mass ($\kappa=\sqrt{16\pi G_N}$),
$\Lambda$ is a bulk cosmological constant defined
so as to have the same dimensions as a vacuum energy density
in $\call_{\rm fields}$,\footnote{In the usual notation,
the cosmological constant term would be written as
${1\over \what\kappa^2}\int d^4\,x d^\del y \sqrt{-\what g}\,\Lambda^*$.
$\Lambda^*$ would have dimension of mass-squared. The relation
is $\Lambda^*=\kappa^2 \Lambda$.}
$\what g_{\what\mu\what\nu}$ ($\what\mu,\what\nu=0,1,2,3,\ldots,3+\del$) 
refers to the full metric tensor in $4+\del$ dimensions,
$\what g_{4\,\mu\nu}$ refers to the $\mu,\nu=0,1,2,3$ part
of $\what g$,  and we have restricted
all matter fields to lie on a brane at $y=0$ following
the prescription of \cite{hanetal} and \cite{wellsetal}.
(We will employ notation and definitions consistent with
\cite{hanetal}.) 
In the linearized approximation, we expand
the metric tensor about the flat-space limit writing
$\what g_{\what \mu\what\nu}=\eta_{\what\mu\what\nu}+\what\kappa
\what h_{\what\mu\what\nu}$ with 
\beq
\what h_{\what\mu\what\nu}=V_\del^{-1/2}\left(
\begin{array}{cc} h_{\mu\nu}+\eta_{\mu\nu}\phi & A_{\mu i} \\
A_{\nu j} & 2\phi_{ij}\end{array}\right)\,,
\label{linexp}
\eeq
where $\phi\equiv \sum_i\phi_{ii}$.\footnote{We note 
that the expansion parameter multiplying the fields is 
$\what\kappa V_\del^{-1/2}=\kappa\equiv
\sqrt{16\pi}/\mplanck$, where $\mplanck$
is the usual Planck mass for three spatial dimensions. Thus, for this 
expansion to be appropriate, it is important that 
the relevant values for the $h_{\mu\nu}$, $A_{\mu j}$ and $\phi_{ij}$
fields all be small relative to $\mplanck$. This
will indeed be the case so long as the string scale, $M_S$,
above which the effective theory that we will be discussing is invalid,
is much smaller than $\mplanck$. This expansion also
assumes that in the absence of quantum excitations a flat metric
is a consistent solution of the equations of motion, 
at least locally on the brane.
This, in turn requires that the vacuum energy density on the brane
be very small (small effective cosmological constant). As discussed
later, this latter can be arranged by an appropriate choice
of the bulk $\Lambda$ in the case that will be of interest
to us where the other Lagrangian terms lead to a very substantial
vacuum energy density on the brane.}
Retaining the leading terms in $\what\kappa$,
the resulting equations of motion take the form:
\beq
R_{\what\mu\what\nu}-{1\over 2}
\eta_{\what\mu\what\nu} R=\what\kappa^2
\eta^\mu_{\what\mu}\eta^\nu_{\what\nu}T_{\mu\nu}\del^\del(y)
-\what\kappa^2 V_\del^{-1}\Lambda  \eta_{\what\mu\what\nu}\,.
\label{eomgr}
\eeq
In Eq.~(\ref{eomgr}),
$T_{\mu\nu}=\left(-
\what g_{\mu\nu}\call+2{\del \call\over \del 
\what g^{\mu\nu}}\right)_{\what g_{\mu\nu}=\eta_{\mu\nu}}$
and $R_{\what\mu\what\nu}-{1\over 2}\eta_{\what\mu\what\nu}R
\equiv {\cal G}_{\what\mu\what\nu}$ with 
\beq
{\cal G}_{\what\mu\what\nu}={\what\kappa\over 2}
\left[\partial^{\what \rho}\partial_{\what\rho}
\what h_{\what\mu\what\nu}-\partial^{\what \rho}\partial_{\what \mu}
\what h_{\what \rho\what\nu} -\partial^{\what\rho}\partial_{\what\nu}
\what h_{\what \rho\what\mu}+\partial_{\what\mu}\partial_{\what\nu}
\what h_{\what \rho}^{\what \rho}-\eta_{\what\mu\what\nu}
\partial^{\what\rho}\partial_{\what \rho}\what h_{\what\sigma}^{\what\sigma}
+\eta_{\what\mu\what\nu}\partial^{\what\rho}\partial^{\what\sigma}
\what h_{\what\rho\what\sigma}\right]\,.
\label{gdef}
\eeq
The next step is to expand:
\beq
h_{\mu\nu}(x,y)=\sum_{\vec n} h_{\mu\nu}^{\vec n}(x){\rm exp}
\left(i {2\pi\vec n\cdot\vec y\over \ldel }\right)\,,\quad
\phi_{ij}(x,y)=\sum_{\vec n}\phi_{ij}^{\vec n}(x) {\rm exp}
\left(i {2\pi\vec n\cdot\vec y\over \ldel }\right)\,,
\label{expansion}
\eeq
and similarly for $A_{\mu i}(x,y)$. Here, the $\vec n\neq0$ modes are the KK
states and we have assumed that all compactification radii are the same.
The result of this expansion yields an effective
Lagrangian and corresponding
equations of motion for the KK modes and Standard Model fields on the brane.
For $\vec n\neq0$, one rewrites $h^{\vec n}_{\mu\nu}$, 
$A^{\vec n}_{\mu\,i}$ and $\phi^{\vec n}_{ij}$
in terms of the physical fields $\htil^{\vec n}_{\mu\nu}$, 
$\widetilde A^{\vec n}_{\mu\,i}$ and $\phitil^{\vec n}_{ij}$
as in Refs.~\cite{hanetal} and \cite{wellsetal}.  
 
Our main focus will be on the Higgs
sector of the theory. We begin
with an arbitrary Higgs potential $V$ that is a function of
a certain set of real Higgs scalar fields $\Phi_i$.
The corresponding contribution to the energy
momentum tensor is $T_{\mu\nu}=\eta_{\mu\nu}V$.
We also have the massive KK states, the Lagrangian mass terms for which are
simply 
\beq
\call_{\rm mass}^{\vec n}=\half m_n^2\left(-\htil^{\mu\nu,\vec n}
\htil_{\mu\nu}^{-\vec n}+\htil^{\vec n}\htil^{-\vec n}\right)
-m_n^2\sum_{i,j=1}^\del\phitil_{ij}^{\vec n}
\phitil_{ij}^{-\vec n}\,,
\label{lmass}
\eeq
where we have singled out fields associated
with a single $\vec n=(n_1,n_2,\ldots,n_\del)$ mode and the
complex conjugate $-\vec n$ fields. (This means that
the total $\call$ is obtained by summing $\call^{\vec n}$
only over $\vec n$ values such that the first non-zero $n_k$
is positive. We denote this restricted sum by $\sum_{\vec n>0}$.) 
In Eq.~(\ref{lmass}),
$m_n^2={4\pi^2 n^2\over \ldel ^2}$ with $n^2={\vec n}^2$
and $\htil=\htil_\mu^\mu$, which is non-zero in the general off-shell
situation.\footnote{Note that we do not agree with
the notation in Eq.~(24) of \cite{hanetal} which implies that the 2nd term
in Eq.~(\ref{lmass}) should only be summed over the  
$\del(\del+1)/2$ independent values of $i,j$. We also see that
$\call^{\vec n}$ of their Eq.~(24) should only be summed over $\vec n>0$,
as specified above.}

The next key ingredient is the coupling between the graviton KK states
and the scalar field contributions to $T_{\mu\nu}$ following from the
last term in Eq.~(\ref{action}). To lowest order in 
$\kappa=\sqrt{16\pi/\mplanck^2}$, and singling out fields
with index $\vec n$ and their complex conjugates, this takes 
the form\footnote{Here, the full $\call$ is again obtained
as $\call=\sum_{\vec n>0}\call^{\vec n}$.}
\bea
{\cal L}_{\rm mix}^{\vec n}
&=&-{\kappa\over2}\left[\left(\htil^{\mu\nu,\vec n}+
\htil^{\mu\nu,-\vec n}\right)T_{\mu\nu}
+\omega_\del\left(\phitil^{\vec n}+\phitil^{-\vec n}\right)T_\mu^\mu\right]\,,
\label{lmix}
\eea
where 
\beq
\omega_\del=\left[2\over 3(\del+2)\right]^{1/2}
\label{odeldef}
\eeq 
and $\phitil=\sum_i\phitil_{ii}$. 

In employing both Eq.~(\ref{lmass}) and (\ref{lmix}), 
we must keep in mind the $\del$ constraints 
$n_i\phitil_{ij}=0$, which means that only $\del(\del-1)/2$
of the  $\phitil_{ij}$ are independent. This is particularly
crucial as we consider
the effects of the mixing between the $\htil,\phitil$
fields and the Higgs field through $\lmix$.  
We first wish to consider whether the mixing term can lead to
non-zero vacuum expectation values for the $\Phi$ Higgs field
and the $\htil,\phitil$ fields. To this end, we look
for an extremum of $\vtot=
V(\Phi_i)-\call_{\rm mass}-\lmix$, where $V(\Phi_i)$ is the relevant part
of $-\call_{\Phi}$, $\call_{\Phi}$ being the Lagrangian for the Higgs field(s).
We will argue later that in general $V(\Phi_i)$ should be the full
effective potential as computed for the Higgs sector. Writing
$\htil^{\mu\nu,\vec n}=\Re(\htil^{\mu\nu,\vec n})+i\Im(\htil^{\mu\nu,\vec n})$
and
$\htil^{\mu\nu,-\vec n}=
\Re(\htil^{\mu\nu,\vec n})-i\Im(\htil^{\mu\nu,\vec n})$,
and similarly for the  $\phitil_{ij}$ fields,
we find that the extremum conditions for  
$\htil^{\mu\nu}$ and $\phitil_{ij}$  give
\beq
\Re(\htil_{\mu\nu}^{\vec n})= \eta_{\mu\nu}
{\kappa\over 3 m_n^2}V
\quad
\Re(\phitil_{ij}^{\vec n}) = -P_{ij} 
{2\kappa\omega_\del\over m_n^2}V, 
\label{extr}
\eeq
where $\eta_{\mu\nu}={\rm diag}(1,-1,-1,-1)$,
$P_{ij}=\del_{ij}-n_in_j/n^2$ and all imaginary components are zero. 
Substituting into $\lmix$ and $\call_{\rm mass}$ gives a result for $\vtot$
which we denote by
$\vtotbar$:\footnote{We note that this
same result is obtained if one computes the $V^2$ terms using
virtual exchanges of the $\htil_{\mu\nu}^{\vec n}$
and $\phitil_{ij}^{\vec n}$ fields at zero momentum. (We believe
that the propagator for $\htil_{\mu\nu}^{\vec n}$ given in \cite{hanetal}
is incorrect. One should use $\Delta^{\htil}_{{\mu\nu,\vec n},
{\rho\sigma,\vec m}}=i\del_{\vec n,-\vec m}B_{\mu\nu,\rho\sigma}
(k^2-m_n^2+i\epsilon)^{-1}$, i.e. no factor of 1/2 in their normalization.
This is required also for consistency with \cite{wellsetal}.)
Indeed, one obtains from virtual $\htil$ and virtual $\phitil$
exchanges the results
$
\sum_{{\rm all}\,\vec n}{i\over -m_n^2}{(i\kappa)^2\over 2}\left(
T^{\mu\nu}T_{\mu\nu}-{1\over 3}T_\mu^\mu T_\nu^\nu\right)\,,$ and
$
\sum_{{\rm all}\,\vec n}{i\over -m_n^2}{(i\kappa)^2\omega_\del^2(\del-1)\over
4} T_\mu^\mu T_\nu^\nu\,,
$
respectively, where we will substitute $T_{\mu\nu}=\eta_{\mu\nu}V$.
To convert to an effective Lagrangian, we must remove the $i$
and multiply by $1/2$ in order to avoid over counting contractions.
After changing the sign in order to convert to an effective potential,
using $\sum_{\vec n>0}$ and writing $T_{\mu\nu}=\eta_{\mu\nu}V$,
we reproduce the corresponding terms of Eq.~(\ref{vtot}).
One can easily see that this must be the result by integrating
out the $\htil$ and $\phitil$ fields in the path integral formulation.
}
\beq
\vtotbar=V
+\sum_{\vec n>0}{1\over m_n^2}{2\kappa^2\over 3}V^2
-\sum_{\vec n>0}{1\over m_n^2}{4\kappa^2}\left({2\over 3}
\,{\del-1\over\del+2}\right)V^2
=V-2\kappa^2V^2{\del-2\over\del+2}\sum_{\vec n>0}{1\over m_n^2}\,,
\label{vtot}
\eeq
where we use the shorthand notation $V$ for $V(\Phi_i)$.
It will be convenient to define 
\beq
\sum_{{\rm all}\,\vec n}{1\over m_n^2}=2\sum_{\vec n>0}{1\over m_n^2}\equiv 
{\dbar\over \kappa^2}{\del +2 \over \del-2}\,,
\label{dbardef}
\eeq
so that we have
\beq
\vtotbar=V-\dbar V^2\,.
\label{leff3}
\eeq

One must next search for an extremum with respect to the
$\Phi_i$. Using the result of Eq.~(\ref{leff3}), one finds
\beq
{\partial \vtotbar\over\partial\Phi_i}
= {\partial V\over\partial\Phi_i}\left[1-2V{\dbar}\right]\,.
\label{partial2}
\eeq
Requiring this to be 0, we obtain:\footnote{We note
that we end up with exactly the same conditions whether we substitute
using Eq.~(\ref{extr}) and then minimize (as we have done) or
take the derivatives of $V(\Phi)-\call_{\rm mass}-\call_{\rm mix}$
with respect to the $\htil_{\mu\nu}^{\vec n}$, $\phitil^{\vec n}_{ij}$ 
and $\Phi$
fields independently and then afterwards substitute. Technically
speaking, the former is appropriate if we are integrating out
the $\htil$ and $\phitil$ fields, whereas the latter is appropriate
if we retain them as physical degrees of freedom. One can
equally well mix the two approaches and obtain the same conditions.
This `matching' 
is important as it guarantees that there is no sensitivity to
exactly where we place the boundary between heavy and light fields.}  
\beq
V={1\over 2\dbar}\,, ~~~~
{\rm or}~~~~{\partial V\over\partial \Phi_i}=0,
~~\mbox{for all}~i\,.
\label{v0val}
\eeq
The 2nd solution corresponds to the usual minimum while the first
is of a very unusual nature, as we shall explore. Whichever
extremum is appropriate, we denote the extremum values of $V$ and $\Phi_i$ by
$V_0$ and $v_i$, respectively. We also denote the 
value of $V$ at the usual minimum by $\vnorm$.

In order to determine which extremum corresponds to the smallest $\vtotbar$,
we compute
\beq
\vtotbar\left(V={1\over2\dbar}\right)-\vtotbar(V=\vnorm)={1\over 4\dbar}-
\left(\vnorm-\dbar\vnorm^2\right)=\dbar\left(\vnorm-{1\over 2\dbar}\right)^2\,.
\label{condition}
\eeq
We see that the $V={1\over 2\dbar}$ extremum is preferred if 
$\dbar<0$, whereas the standard extremum is preferred for $\dbar>0$ unless
$\vnorm={1\over 2\dbar}$.
The procedure of Ref.~\cite{hanetal} yields  
$\dbar={2\over M_S^4(\del-2)}$, where $M_S$ is an ultraviolet cutoff.
This suggests that $\dbar>0$.  However, the ultraviolet cutoff is
the point at which the physics of the string enters. The exact manner
in which the divergent sum is regularized is thus uncertain and either
sign for $\dbar$ is possible, as considered, for example, 
in Ref.~\cite{hewett}.\footnote{A simple example is $\zeta$ regularization.
Defining $\zeta(x)\equiv \sum_{n=1}^\infty \,n^{-x}$, 
$\zeta(x)$ is easily computed and is positive
for $\Re(x)>1$. But, using analytical extension to define the divergent
summation for negative $x$ gives $\zeta(-1)=-1/12,\zeta(-5)=-1/152,\ldots$. 
In general, $\zeta(1-2m)$ has sign $(-1)^m$, i.e. a negative
value results for odd integer $m$ even though $\zeta(1-2n)$ is
formally a sum of positive numbers. In fact, the large $n$ 
terms in the summation
$\sum_{\vec n}{1\over m_n^2}$ behave like $\sum_n c_\del n^{\del-3}$
with $c_\delta=\frac{\ldel ^2\pi^{\delta/2-2}}{2\Gamma(\delta/2)}$. Thus,
we could write $\sum_{\vec n}{1\over m_n^2}=c_\del 
\zeta(\del-3)+d_\del$, where $d_\del$ is a finite correction. However, 
since $\zeta$ regularization is not the only possibility,  
results obtained in this specific manner would probably be misleading.
The only firm conclusion is that string physics could
act to regulate apparently divergent sums in a manner
such that $\dbar<0$.}
Thus, we consider $\dbar$ to simply be a parameter determined
by the detailed physics at the string scale.

Given a definite minimum, we must consider an appropriate quantum state 
expansion.  To this end, we first note that for a set of scalar fields
$\Phi_i$
one has $T_\mu^\mu=-2\call_\Phi+2V(\Phi_i)$, where
$\call_\Phi=\sum_i \call_{\rm kin,\, i} - V(\Phi_i)$,
where $\call_{\rm kin,\, i}=\half (\partial_\rho \Phi_i)
(\partial^\rho \Phi_i)$. 
By substituting the
vacuum field values of Eq.~(\ref{extr}) into Eq.~(\ref{lmix}) 
and summing over all $\vec n>0$, we find
\beq
\lmix\sim\kappa^2V_0{\del-2\over\del+2}\sum_{\vec n>0}{1\over m_n^2}
T_\mu^\mu=\half \dbar V_0T_\mu^\mu=-\dbar V_0
\left[\call_\Phi-V(\Phi_i)\right]\,.
\label{lmixmin}
\eeq
To determine the appropriate quantum expansion
for the scalar fields, we focus on the derivative terms
\beq
\call_\Phi+\lmix\ni \left(1-\dbar V_0\right)\half
\sum_i\partial^\rho\Phi_i\partial_\rho\Phi_i
\,.
\label{kemix}
\eeq
(For later reference, we note that $\call_\Phi+\lmix$ also contains
non-derivative terms of the form $[2\dbar V_0-1]V(\Phi_i)$.)
From Eq.~(\ref{kemix}) we see that, for the $V_0={1\over 2\dbar}$ minimum, half
of each usual
$\call_{\rm kin,\,i}$ derivative term is canceled by $\lmix$;
whatever the value of $V_0$, 
we must rescale the $\Phi_i$ in order to have canonical
normalization for their kinetic energy terms. We write
\beq
\Phi_i=\what\Phi_i\left(1-\dbar V_0\right)^{-1/2}\,.
\label{phirescale}\eeq
Note that rescaling is not necessary for the KK $\htil$
and $\phitil$ fields since $\lmix$ does not
involve their derivatives.

The next step is to expand $\vtot= V(\Phi_i)-\call_{\rm mass}
-\lmix$ 
about the extremum. We write 
\beq
\what\Phi_i=\what v_i+{s_i}\,,
\quad \Re(\phitil^{\vec n}_{ij})=-P_{ij}{2\kappa\omega_\del\over m_n^2}V_0
+\sum_s e_{ij}^s{s^s_{\vec n}\over \sqrt 2}\,.
\label{expan}
\eeq
It will not be necessary to expand $h_{\mu\nu}^{\vec n}$
about its minimum since $T^{\mu\nu}\propto \eta^{\mu\nu}$
and the spin-2 quantum states have
polarizations $\eps_{\mu\nu}^{\vec n}$ such that 
$\eta^{\mu\nu}\eps_{\mu\nu}^{\vec n}=0$.
We may also choose to define the $s^s_{\vec n}$ states so that only
$e_{ii}^{s=1}\neq 0$; $e_{ii}^1=(\del-1)^{1/2}$ is required for
correct normalization (summation over $i$ is implied in both formulae).
Only $s_{\vec n}^1$ has the potential for mixing with the $s_i$ states.
The resulting form for $\vtot$ is:
\bea
\vtot&=&V_0-\dbar V_0^2+{1\over 2}\sum_{\vec n>0}m_n^2 (s^1_{\vec n})^2
+2\sqrt 2\kappa
\omega_\del(\del-1)^{1/2} \sum_{\vec n>0} 
s^1_{\vec n}\sum_js_j\left({\partial V\over\partial \what\Phi_j}
\right)_{\what\Phi_k=\what v_k,~{\rm all}~k}\nonumber\\
&\phantom{=}&+(1-2\dbar V_0)
\sum_{ij}s_is_j\left({1\over 2}{\partial^2 V\over\partial\what
\Phi_i\partial\what\Phi_j}
\right)_{\what\Phi_k=\what v_k,~{\rm all}~k}
+\ldots\,,
\label{vtotform}
\eea
where ${\partial V\over\partial \what\Phi_i}
=(1-\dbar V_0)^{-1/2}{\partial V\over \partial\Phi_i}$.
The physics at the two different extrema of Eq.~(\ref{v0val})
is quite different.  If ${\partial V\over\partial\hat\Phi_i}=0$ for all $i$,
then Eq.~(\ref{vtotform}) shows that there is no tree-level mixing
between $s_{\vec n}^1$ and any of the $s_i$.\footnote{As
we shall show shortly, the same is true at the one-loop
level.}  The Higgs and KK modes
remain in separate sectors. The case of $V_0={1\over 2\dbar}$ is more subtle.
To illustrate, we assume that there is only one $\Phi_i$. In this case,
the mass terms for the quantum fluctuations read
\beq
\vtot\rightarrow\half\sum_{\vec n>0}\left[m_n^2 (s_{\vec n}^1)^2+2\eps
s_{\vec n}^1\,s\right]\,,~~~\mbox{with}~~~\eps\equiv 2\sqrt 2\kappa
\omega_\del (\del-1)^{1/2}
\left({\partial V\over\partial \what\Phi}\right)_{\what\Phi=\what v}\,,
\label{vtot1}
\eeq
where Eq.~(\ref{phirescale}) implies that 
${\partial V\over \partial\what\Phi}=\sqrt 2 
{\partial V\over \partial \Phi}$ when $V_0={1\over 2\dbar}$.
In order to determine whether or not we are in a local minimum,
we must diagonalize $\vtot$.
The mass matrix takes the form
\beq
{\cal M}^2=\left( \begin{array}{c c c c c}
0 & \rho_1 \eps & \rho_2\eps & \rho_3\eps & \ldots \\
\rho_1\eps & \rho_1m_1^2 & 0 & 0 & \ldots \\
\rho_2\eps & 0 & \rho_2m_2^2 & 0 & \ldots \\
\rho_3\eps & 0 & 0 & \rho_3m_3^2 & \ldots \\
\ldots & 0 & 0 & 0 & \ldots \\
\end{array}\right)
\eeq
where the notation $\rho_n\eps$ and $\rho_nm_n^2$ means
that there are actually $\rho_n=\sum_{\vec n>0,\,\vec n^2=n^2}$ 
identical entries in the matrix.
The precise values of $\rho_{1,2,3,\ldots}$ depend upon
$\del$. For example, $\rho_1=\del$. (The $\vec n>0$ solutions
of $1^2=n_1^2+n_2^2+\ldots+n_\del^2$ are $n_i=1$, $n_{k\neq i}=0$
for any $i=1,\ldots,\del$.) A typical value of $m_n^2={4\pi^2n^2\over \ldel ^2}$ 
is set by the relation 
${4\pi^2\over \ldel ^2}=\left({\mplanck\over \sqrt{8\pi}}
\right)^{-4/\del}M_S^{2+4/\del}$
from which one obtains ${4\pi^2\over \ldel ^2}=1.7\cdot 10^{-7}~\mbox{eV}^2$,
$3\cdot 10^3~\mbox{eV}^2$, $4\cdot 10^8~\mbox{eV}^2$, 
$5\cdot 10^{11}~\mbox{eV}^2$, $6\cdot 10^{13}~\mbox{eV}^2$,
$2\cdot 10^{16}~\mbox{eV}^2$
for $\del=2,3,4,5,6,8$ for a $4+\del$
dimension Planck mass $M_S$ of order $1\tev$. 
This is to be compared to the off-diagonal entries
characterized by $\eps\sim m^2 v/\mplanck\leq 10^7~\mbox{eV}^2$ 
for $m<1\tev$.
Thus, for $\del\geq 4$ the off-diagonal entries are always
small compared to the diagonal entries. Even more importantly,
the upper cutoff in $m_n^2$ to which we sum and
which dominates the relevant summations is of order $M_S^2
\sim 10^{24}~\mbox{eV}^2$ (for $M_S\sim 1\tev$)
or larger, which is much larger than $\eps$.
Thus, for the relevant matrix entries, the 
$s_{\vec n}^1$ KK states mix slightly with $s$ with a mixing angle 
$\theta_{\vec n}\sim -\eps {1 \over m_n^2}$.
The physical eigenstate corresponding to the original $s$ is rotated to 
\beq
s_{\rm phys}\simeq \left(\begin{array}{c}
 \Pi_n c_{\theta_n} \\ s_{\theta_1} \\ s_{\theta_2}\\
\ldots \end{array}\right)\simeq
\left(\begin{array}{c} 1\\ \theta_1 \\ \theta_2 \\ \ldots \end{array} 
\right)\simeq
\left(\begin{array}{c} 1 \\ -\rho_1\eps/m_1^2 \\ 
-\rho_2\eps/m_2^2 \\ \ldots \end{array}
\right)\,,
\label{sphys}
\eeq
where we use the short-hand notation of lumping all $\rho_n$
of the states of a given $m_n$ together.
The mass of $s_{\rm phys}$ is given by\footnote{It is remarkable
that this result is obtained whether we treat the $\phitil^{\vec n}_{ij}$
fields as being light, i.e. without integrating them out, (as we have done) 
or first integrate them out. Just as for the minimization condition
of Eq.~(\ref{v0val}), the final expression
for $\msphys^2$ does not depend upon where the dividing
line between `heavy' and `light' fields is placed.}
\beq
\msphys^2=-\eps^2\sum_{n}{\rho_n\over m_n^2}\to
-\eps^2\sum_{\vec n>0} {1\over m_n^2}
\sim -{8\over 3}\dbar {\del-1\over\del-2}
\left({\partial V\over\partial\what\Phi}\right)_{\what\Phi=\what v}^2
\,,
\label{msphys}
\eeq 
where we have used Eqs.~(\ref{dbardef}) and (\ref{odeldef}).
We note that the $\dbar<0$ requirement, needed to ensure
that we are expanding about a local minimum that is deeper
than the standard minimum, is also that which implies
a positive mass-squared for $s_{\rm phys}$.
Mixing of $s$ with
the full tower of KK states and whatever physics is present
at the string scale to cutoff the ultraviolet divergence of 
$\sum_{\vec n}{1\over m_n^2}$ is critical to obtaining $\msphys^2>0$.

Let us now make a few comments before turning to phenomenology.
First, it is very amusing to note that the $\dbar<0$ minimum
yields nonzero $v$ at the minimum even if $V(\Phi)$ itself does not
have a minimum with $\Phi\neq0$.
In particular, 
\beq
V(\Phi)=\half m^2 \Phi^2+\Xi
\label{vform}
\eeq
is entirely satisfactory,
provided $V_0=\half m^2v^2+\Xi={1\over 2\dbar}<0$, where 
$v=\sqrt 2\hat v$.
For this form of $V(\Phi)$ we have 
\beq
\msphys^2=-{32\over 3}\dbar {\del-1\over\del-2}m^4\what v^2\,,
\label{msphys2}
\eeq
which would be of order $\what v^2$ if $m\sim |\dbar|^{-1/4}\sim M_S$, as might
possibly be natural. We note that if $m$ is of this size, then
to achieve $V_0={1\over 2\dbar}$ it is necessary that $\Xi$
be negative with absolute magnitude of order $M_S^4$. This is all
that is required in order for the interactions of the tower
of KK states with the Higgs field to generate electroweak symmetry breaking.

Second, we argue that in our approach it is incorrect
to include in $V(\Phi)$ tree-level diagrams containing
virtual exchanges of the $s_{\vec n}$ fields. (These would, in particular,
create effective $\Phi^4$ and higher interactions for 
$V(\Phi)=\half m^2\Phi^2+\Xi$).
This is because
the graphs required are one-particle reducible (in the $s_{\vec n}^1$ fields).
However, true one-loop corrections to $V(\Phi)$ should be included
beyond the tree level.
The usual Higgs, fermion and vector loops will be discussed
later.  In addition, there are one-loop
corrections that first appear at ${\cal O}\left(\kappa^2\right)$.
For example, the sea-gull $\phitil\phitil \Phi\Phi$ vertex of order
$\kappa^2$ gives rise to a one-loop diagram coupling two pairs of $\Phi$'s
together via a loop containing two (independent) $\phitil$ exchanges.
Summing over all the independent $\vec n$ states for each $\phitil$
and integrating over the loop momentum, one obtains a one-loop
$\Phi^4$ interaction with coefficient
of order $[\kappa^2]^2[\dbar/\kappa^2]^2/\dbar\sim \dbar$.
This is of similar form to the $\dbar V^2$ term in Eq.~(\ref{leff3}),
but suppressed by a one-loop numerical factor.
We will consistently neglect all  diagrams containing KK loops.

Thirdly, we wish to note that the actual size of $\dbar$
is extremely uncertain.
For example, the recent work of Ref.~\cite{cutoff} suggests that
the brane recoil effects will provide an effective cutoff
for the $\sum_{\vec n}{1\over m_n^2}$ that can reduce the size of $\dbar$:
$\dbar\to\dbar\times \left({f^2\over M_S}\right)^{\del-2}$ for $\del>2$,
where $f$ is the brane tension.
However, if $\dbar<0$ and $V_0=1/(2\dbar)$, this does not affect
the size of $\dbar V_0$. As a result, if $\dbar<0$ 
its exact magnitude is not critical to the resulting phenomenology.

Finally, we note that the result of Eq.~(\ref{msphys2})
for the physical Higgs mass squared will receive loop corrections
from a number of sources. For example, there are the
corrections to the Higgs mass coming from the virtual KK
corrections as computed in Ref.~\cite{hanetal}.
These are proportional to the $m^2$ parameter appearing
in $V$ times functions of
$m^2/M_S^2$.  Most naturally, all mass parameters are of order 
$|\dbar|^{-1/4}\sim M_S$,
in which case the virtual KK correction would be of the order
of $M_S^2$, and therefore possibly somewhat larger than $\msphys^2$
(which might be of order $v^2$). More importantly, 
there are one-loop contributions from vector boson and fermion loops.
As usual, these are quadratically divergent unless we introduce
the usual supersymmetric partners.  
In this paper, we do not consider the supersymmetric
extension. Quadratically divergent contributions
are then absorbed in the usual renormalization procedure.

\section{Electroweak Phenomenology}
\medskip

Let us first focus on vector boson mass generation.
There are two contributions; one coming from $\lmix$ and
the other from the standard $\call_{\rm kin}$ kinetic energy
portion of the scalar Lagrangian. After substituting the vacuum
values for $\Re(\htil^{\mu\nu,\,\vec n})$ and $\Re(\phitil_{ij}^{\vec n})$
given in Eq.~(\ref{extr}) into Eq.~(\ref{lmix}), we find
the result of Eq~(\ref{kemix}) for the pure derivative terms. 
After rescaling to $\what\Phi$, one obtains the usual canonically
normalized derivative form 
$\lmix+\call_{\rm kin}\ni
\half \partial_\rho\what\Phi \partial^\rho\what\Phi$. At this point, we
generalize to real doublet notation and expand to the usual gauge-invariant
derivative form to obtain (regardless of the value of $V_0$)
\beq
\lmix+\call_{\rm kin}\ni \half\left(D^\rho H\right)^\dagger
\left(D_\rho H\right)\,,
\label{canonicalform}
\eeq
 where for the simple one-doublet model
we  employ $H=(0,\what\Phi)$ and $D_\rho=\partial_\rho+igA_\rho^aT^a$
with $T^a=\tau^a/2$.    Keeping
only the $W$ and $Z$ boson portions of Eq.~(\ref{canonicalform}) and
writing $\what\Phi=\what v+s$ as before, we find
\beq
\lmix+\call_{\rm kin}\ni
{1\over4}g^2(\what v+s)^2W_\rho^+W^{-\,\rho}
+{1\over 8}\left(g^2+g^{\prime\,2}\right)(\what v+s)^2Z_\rho Z^\rho\,.
\label{wzlmass}
\eeq
However, before proceeding further, we must demonstrate
that the $W$ and $Z$ fields themselves do not need to be rescaled
in order to have canonical normalization.  The crucial result is that
following from Eq.~(\ref{lmixmin}):
\beq
\lmix=\half\dbar V_0 T_\mu^\mu
\label{lmixspecial}
\eeq 
after substituting
the vacuum values of the $\htil$ and $\phitil$ fields.
From the form of $T_{\mu\nu}^V$ for a vector field given in Eq.~(37)
of \cite{hanetal}, one finds that the $F^{\rho\sigma}F_{\rho\sigma}$
kinetic energy part of $T_\mu^{V\,\mu}$ is zero.

Thus, we can proceed to read off masses and couplings from Eq.~(\ref{wzlmass}).
One finds $m_W^2={1\over 4}g^2\what v^2$ and 
$m_Z^2={1\over 4}(g^2+g^{\prime\,2})\what v^2$, which are the
usual expressions but 
in terms of the vev of the rescaled field, $\what \Phi$.
(Note that the standard result for the ratio $m_W/m_Z$ is preserved.)

We must now compute the coupling of $WW$ and $ZZ$ to $\sphys$.
We focus on the $WW$ coupling, that for $ZZ$ being entirely analogous.
In principle, both $WWs$ and $WWs_{\vec n}^1$ couplings contribute
after mixing. The former coupling is easily read off
from Eq.~(\ref{wzlmass}). The latter coupling is obtained by substituting
the kinetic part of the Higgs $T_{\mu\nu}$ into Eq.~(\ref{lmix}),
yielding a $\lmix^{\vec n}$ term of the form 
$\sqrt 2\kappa \omega_\del(\del-1)^{1/2}s_{\vec n}^1\call_{\rm kin}$.
The net result for the $WWs_{\rm phys}$ coupling is
\beq
{g^2 \what v\over 2}+\sum_{\vec n>0}\kappa\omega_\del (\del-1)^{1/2}
{g^2\what v^2\over 2\sqrt 2}\theta_{\vec n}={g m_W}
-{1\over 3}\dbar {\del-1\over\del-2}
g^2\what v^2\left({\partial V\over\partial\what \Phi}
\right)_{\what\Phi=\what v}\,.
\label{wwnet}
\eeq
The 2nd term only contributes for the $\dbar<0$ minimum where
the mixing angles, proportional to  $\partial V/\partial\what\Phi$
at the minimum, are non zero.  In this case, assuming the simple
$V(\Phi)$ form of Eq.~(\ref{vform}), this 
2nd term is of order $gm_W(\what v^2m^2/M_S^4)$ and will yield a small
correction to the usual $gm_W$ strength for the $WW\sphys$ coupling.
To the extent that it can be neglected, $WW$ scattering
(for example) will not violate unitarity bounds
if $\msphys\lsim 1\tev$.
The unitarity problems associated with the correction term
in Eq.~(\ref{wwnet}) would, in fact, generally
be smaller than those arising from 
virtual $\htil$ and $\phitil$ contributions to $WW$ scattering. These latter
are summarized in the amplitude form
\beq
{\cal A}(s)=-{\kappa^2\over 2}\sum_{{\rm all}\,\vec n}
{1\over s-m_n^2}{\cal T}\,,
\label{calaform}
\eeq
where ${\cal T}=T_{\mu\nu}T^{\mu\nu}-{1\over \del+2}T_\mu^\mu T_\nu^\nu$, 
where for $WW$ scattering the $T_{\mu\nu}$
would be that for the $W$'s. As usual, 
for $s<M_S^2$,  $\sum_{\vec n>0}{1\over s-m_n^2}$
would be dominated by contributions at the $M_S$ scale
yielding $-\kappa^2\sum_{\vec n>0}{1\over s-m_n^2}\sim \dbar\sim 1/M_S^4$,
implying an effective contact interaction form.  The $WW$ unitarity
problems deriving from the correction term of Eq.~(\ref{wwnet})
and from the effective contact interaction
of Eq.~(\ref{calaform}) can both be suppressed simply
by taking $M_S$ to be large. Still, it would be interesting to
try to probe the correction term of Eq.~(\ref{wwnet}) by a high
precision measurement of the Higgs-$WW$ coupling once the Higgs boson
has been discovered.

Let us now analyze the fermion sector.
We again use the general result of Eq.~(\ref{lmixspecial}).
The rescaling of the fermion
fields is determined by noting that for the fermionic $T_{\mu\nu}^{\rm F}$
of Eq.~(42) of Ref.~\cite{hanetal} we have
\beq
T_\mu^{F\,\mu}\ni -3\anti\psi i\gamma^\rho D_\rho \psi\,,
\label{tmumuf}
\eeq
where we have dropped total derivative terms.
Thus, we have 
\beq
\call_\psi+\lmix\ni \left(1-{3\over 2}\dbar V_0\right)
\overline\psi i\gamma^\rho D_\rho \psi\,,
\label{fsum}\eeq
implying that we should rescale to 
\beq
\what\psi=\left(1-{3\over 2}\dbar V_0\right)^{1/2}\psi\,.
\label{psirescale}
\eeq
Considering next the Yukawa coupling, for which we use the notation
$\call_\psi\ni -\fy \overline\psi\psi\Phi$, implying 
$T_{\mu\nu}^\psi=\eta_{\mu\nu}\fy \overline\psi\psi\Phi$, we find that
\beq
\call_\psi+\lmix\ni -\left(1-2\dbar V_0\right)\fy \overline\psi\psi\Phi
=-{\left(1-2\dbar V_0\right)\over \left(1-{3\over 2}\dbar V_0\right)
\left(1-\dbar V_0\right)^{1/2}}\fy \overline{\what\psi}\what\psi\what\Phi \,.
\label{fsumfy}
\eeq
If $1-2\dbar V_0\neq 0$, then this simply amounts to a redefinition
of the Yukawa coupling strength $\fy $, which does 
not affect the standard relation between
the $s \what\psi\anti{\what \psi}$ 
coupling and the mass $m_{\what\psi}$ induced
by $\what v$. Any Yukawa coupling term involving the 
real doublet $H$ will always contain
the combination $\what v+s$.  
However, for the $\dbar<0$ minimum,
$1-2\dbar V_0=0$ and it appears that the Yukawa interaction is automatically
zeroed.  

As remarked earlier, if $\dbar>0$ then $\partial V/\partial\Phi=0$
is required at the minimum,
in which case Eq.~(\ref{vtotform}) implies that there
is no mixing between $s_{\vec n}^1$ and the Higgs fields. As we have
stressed, since $V(\Phi)$ is the full effective potential, this is not 
just a tree-level result. Even though there are diagrams involving
Higgs boson, fermion and vector boson loops that 
appear to mix $s_{\vec n}^1$ with $s$, the full
calculation is such that the sum of all such diagrams simply serves
to modify $\call_{\rm mix}$ in such a way that it is the full effective
potential $V$
that should be written in $T_\mu^\mu$, and not just the tree-level potential.
We will explicitly demonstrate this for one-loop. We believe the result
to be general.
The one-loop contribution to the effective 
potential $V$ is conveniently summarized as
\beq
\delta V(\Phi)={1\over 64\pi^2}\left[3m_V^4(\Phi)+m_\Phi^4(\Phi)-
4m_\psi^4(\Phi)\right] \ln {\Phi^2\over M^2}\equiv \delta V_V(\Phi)+
\delta V_\Phi(\Phi)+\delta V_\psi(\Phi)\,,
\label{dveffform}
\eeq
where $M=\vev{\Phi}$. In the above, $m_V(\Phi)=e\Phi$ (where $e$
is a generic gauge coupling), $m_\Phi^2(\Phi)=m_\Phi^2+\half \lambda \Phi^2$
(where we use the Higgs Lagrangian form $\call_\Phi=
\half (\partial_\mu\Phi)^2-V_{\rm tree}(\Phi)$ with 
$V_{\rm tree}(\Phi)=\half m_\Phi^2\Phi^2+{\lambda\over 4!}\Phi^4$)
and $m_\psi(\Phi)=\fy\Phi$ ($\fy$ being the Yukawa coupling
appearing in $\call_{\psi}=\overline\psi i 
\gamma^\rho D_\rho \psi-\fy\overline\psi\psi\Phi$). We wish to demonstrate
that the $\htil$ and $\phitil$ fields interact with
$\delta V(\Phi)$ just as they do with the tree-level $V(\Phi)$, 
see Eq.~(\ref{lmix}). To do so, consider the Lagrangian constructed
by adding $\call_{\rm mix}^{\vec n}$ to $\call_\Phi$. (For convenience,
we focus on a single value of $\vec n$.) Since we are interested
in expanding about the potential minimum
for which $\htil_{\mu\nu}\propto \eta_{\mu\nu}$, 
we may write $\call_{\rm mix}=\alpha T_\mu^\mu$,
where $\alpha\equiv -{\kappa\over 2}
\left(\quarter \eta^{\mu\nu}[\htil_{\mu\nu}^{\vec n}+\htil_{\mu\nu}^{\vec -n}]
+\omega_\del [\phitil^{\vec n}+\phitil^{\vec -n}]\right)$.
Let us first focus on the scalar sector contribution, for which
$T_\mu^\mu=-\partial_\rho\Phi 
\partial^\rho\Phi+4V_{\rm tree}(\Phi)$.
We now recast $\alpha T_\mu^\mu+\call_\Phi$
in the form of the original $\call_\Phi$ by using appropriate rescalings
of fields and couplings.  First, for a canonical kinetic energy normalization
we must define $(1-2\alpha)\half\partial_\rho\Phi\partial^\rho\Phi=\half
\partial_\rho\Phip\partial^\rho\Phip$, implying $\Phip=(1-2\alpha)^{1/2}\Phi
\sim (1-\alpha)\Phi$. (Since we are considering the linearized expansion
in powers of $\kappa$, we need only keep ${\cal O}(\alpha)$ terms.)
One then finds that the net coefficient of $\Phi^{\prime\, 2}$ is
$\sim -\half(1-2\alpha)m_\Phi^2$,
implying that we should define 
$m_{\Phip}^{2}\sim (1-2\alpha)m_{\Phi}^2$. In contrast, we find that
rescaling of $\lambda$ is not necessary; $\lambda^\prime=\lambda$.
We now compute the contribution to the
one-loop potential coming from the scalar sector using the $\Phip$
Lagrangian. The result is analogous to
that contained in Eq.~(\ref{dveffform}), namely 
${1\over 64\pi^2}m_{\Phip}^4\ln{\Phi^{\prime\,2}\over M^{\prime\,2}}$, with
$m_{\Phip}^4(\Phip)=(m_{\Phip}^2+\half\lambda^\prime\Phi^{\prime\,2})^2
\sim (1-4\alpha)m_\Phi^4(\Phi)$ (both terms in $m_{\Phip}^2(\Phip)$
scale in the same way) and $\ln{\Phi^{\prime\,2}\over M^{\prime\,2}}=
\ln {\Phi^2\over M^2}$. The net result is $\call_{\Phi}+\call_{\rm mix}\ni
-(1-4\alpha)V_{\rm tree}(\Phi)-(1-4\alpha)\delta V_\Phi(\Phi)$, 
which is equivalent to including $\delta V_\Phi(\Phi)$
in both $\call_\Phi\ni -V(\Phi)$ and in $\call_{\rm mix}= 
\alpha T_\mu^\mu\ni 4\alpha V(\Phi)$. We turn next to the vector boson loop.
After rescaling, $m_V^4(\Phip)=e^4\Phi^{\prime\,4}\sim (1-4\alpha)e^4\Phi^4$,
which is equivalent to including $\delta V_V(\Phi)$ in the full
$V(\Phi)$ in both $\call_{\Phi}$
and $\call_{\rm mix}$.\footnote{There is no rescaling of the gauge
fields since the gauge kinetic terms do not appear in $T_\mu^\mu$
for the vector theory.  As a result, the gauge coupling to
the $\Phi$ is also
not rescaled since the gauge interactions arise from the covariant
derivative $D_\rho\ni ieA_\rho$.} Finally, consider the fermionic sector.
Since $\call_\psi+\alpha T_\mu^\mu \ni (1-3\alpha)
\overline\psi i\gamma^\rho D_\rho\psi$, canonical normalization for
the net kinetic terms requires rescaling 
$\psi^\prime\sim \left(1-{3\over 2}\alpha\right)\psi$. The scaling
for the Yukawa constant is determined by 
requiring $\call_\psi+\alpha T_\mu^\mu\ni
-(1-4\alpha)\fy\overline\psi\psi\Phi=
-\fy^\prime \overline {\psi^\prime}\psi^\prime\Phip\sim -\fy^\prime
(1-3\alpha)(1-\alpha)
\overline\psi\psi\Phi$, implying $\fy^\prime\sim \fy$. We then have
$m_\psi^4(\Phip)=(\fy^\prime{\Phip})^4\sim (1-4\alpha)(\fy\Phi)^4$, 
equivalent to including $\delta V_\psi(\Phi)$
in $V(\Phi)$ in both $\call_\Phi$ and $\call_{\rm mix}$.

We reemphasize that we have 
consistently neglected all terms that arise from expanding
the metric tensor to higher order in powers of 
$\kappa(h_{\mu\nu}+\eta_{\mu\nu}\phi_{ii})$. These would lead (being
very schematic) to corrections of order $\kappa^2\htil\htil T$ and
$\kappa^2\phitil\phitil T$ to $\lmix$ of Eq.~(\ref{lmix}).
The extremum solution will be shifted by such terms. 
The higher order expansion terms also
lead to new contributions to $m_W$, for instance
through the quartic $WW\Re(\phitil^{\vec n})\Re(\phitil^{\vec m})$
interactions that will contribute to $m_W^2$ when 
the (shifted) vacuum expectation values for the 
$\Re(\phitil^{\vec n})$ are substituted. 

\section{Basic Issues and Phenomenology for \boldmath $V_0\neq 0$}

We first wish to point out that the basic gauge-theory interaction
strengths are not altered by the rescaling required when $V_0\neq 0$.
For example, consider the interaction of the fermionic $\psi$
field with a vector field.
After the rescaling, $\call_\psi+\call_{\rm mix}\ni  
\overline{\what\psi} i\gamma^\rho D_\rho \what\psi$ 
is of canonical form.  This,
in combination with the fact that there is no rescaling for the vector fields
contained in $D_\rho$ implies that the $W\what\psi\anti{\hat \psi}$ and
$Z\what\psi\anti{\what\psi}$ couplings are the same as always. 
The same remarks apply also to the interactions of the Higgs fields
with the vector fields.  Indeed, after rescaling we have already
noted that the Higgs kinetic energy terms have a canonical
normalization; i.e. after rescaling and making
the derivatives covariant $\call_\Phi+\call_{\rm mix}\ni
\sum_i \half (D_\rho\Phi_i)(D^\rho\Phi_i)$.  This structure guarantees
standard gauge couplings for the Higgs bosons. Clearly,
the gauge structure of the theory is being preserved precisely because
the vector fields do not require rescaling (due to the fact, noted earlier,
that their contribution to the trace of the energy momentum tensor
vanishes at tree-level\footnote{However, the anomaly in the trace
of the energy momentum tensor would modify this conclusion; 
see Ref.~\cite{collins}.}).

If $\dbar>0$, then electroweak symmetry
breaking is only possible if $V$ itself has a minimum for non-zero
$\Phi$. A typical form is $V=\lambda(\Phi^2-v^2)^2+\Xi$,
leading to $V_0=\Xi$. The Higgs self interactions induced
by such a potential will not be related in the usual way to
the Higgs mass if $V_0\neq 0$ and the Higgs fields are rescaled.

Another issue regarding the $\dbar>0$ case is the following.
The contribution of $\lmix$ to $\vtot$ means that when
the KK modes take their appropriate values at the minimum
of $\vtot$, $\lmix$ yields a negative contribution to
$\vtot$ as reflected in the form $\vtotbar=V-\dbar V^2$ of Eq.~(\ref{leff3}).
Then, for large enough $V(\Phi)$ (large $\Phi$) $\vtotbar$ is unbounded
from below.  Thus, the standard electroweak symmetry breaking
minimum is intrinsically unstable.  However, if early universe
evolution is such that we enter the $\Phi=v$ minimum, the height
of the potential barrier that must be overcome to reach large $\Phi$
is given by the maximum of $\vtotbar$, i.e. $(\vtotbar)_{\rm max}=1/(4\dbar)$,
which is most naturally of order $\vtotbar\sim M_S^4$.  For large enough $M_S$,
the tunneling probability will be very small. 
From another perspective, the model we are discussing
is only an effective theory valid below a certain cutoff scale (the
string scale $M_S$). In this context, we need only require that
$\vtot(\Phi_i)>V_0-\dbar V_0^2$ for field strengths $\Phi_i$
smaller than $M_S$, as is the case.

Finally, there is the question of how to resolve $V_0\neq 0$
(required to be of order $M_S^4$ in magnitude if $\dbar<0$
and most naturally of this magnitude even if $\dbar>0$) 
with the known fact that the  
vacuum energy on the brane (i.e in our world of three
spatial dimensions) is very small.
We first note that adding an explicit cosmological constant 
that exists only on the brane by including a term in Eq.~(\ref{action})
of the form $\int d^4x \sqrt{-\what g_4}\Lambda_4$ is simply
equivalent to shifting
the value of the constant $\Xi$ as already included in the Higgs potential.
Indeed, after expanding
$\sqrt{-\what g_4}\sim 1+{\kappa\over 2}h+2\kappa \phi$ (where $h=h_\mu^\mu$
and $\phi=\phi_{ii}$, i.e. not $\htil$ and $\phitil$),
we find that the effect is to alter 
Eq.~(\ref{leff3}) to $\vtotbar=(V-\Lambda_4)-\dbar(V-\Lambda_4)^2$. 
Clearly this amounts to a redefinition of $\Xi\to\Xi-\Lambda_4$. 
Further, the effective
vacuum energy seen by the four-dimensional subset of
the basic equations of motion, Eq.~(\ref{eomgr}), including
the massless gravitational
($\vec n=\vec 0$) modes, would be $V_0-\Lambda_4$.  
Thus, if the cosmological constant
only resides on the brane, it is equivalent to shifting 
$\Xi\to \Xi - \Lambda_4$. As a result, the vacuum energy
at the minimum will be exactly as before. For instance, the $\dbar<0$ minimum
would correspond to an effective vacuum energy
of $V_0-\Lambda_4={1\over 2\dbar}$, which is not only non-zero
but has a very large magnitude of order $M_S^4$.

At least one solution to this potential problem, which would also
seem to be quite natural in the string theory context, is to introduce
a cosmological constant throughout the bulk in the manner
of the $\Lambda$ terms of Eqs.~(\ref{eomgr}) and (\ref{action}).
Using Eq.~(\ref{eomgr}),
and multiplying the resulting equation by 
$\exp\left(-i{2\pi \vec n\cdot\vec y\over \ldel }\right)$ 
and integrating over the $y$ coordinates projects
out the equations of motion for the various $\vec n$ components
of the fields.  The crucial point is that the $\Lambda$ term in
Eq.~(\ref{eomgr}) contributes only for $\vec n=\vec 0$.
 The $\vec n\neq \vec 0$ considerations we have been discussing
are unaltered. For $\vec n=\vec 0$, after integration over $y$, 
in the 4-dimensional space the right hand side of Eq.~(\ref{eomgr})
takes the form $\what\kappa^2( T_{\mu\nu}-\Lambda\eta_{\mu\nu})$.
A vacuum energy
density $T_{\mu\nu}=\eta_{\mu\nu}V_0$ can be canceled by choosing
$\Lambda=V_0$.
In the $\dbar<0$ case, with $V_0\sim -M_S^4$, one requires  
$\Lambda<0$ and of order $M_S^4$.\footnote{A complete expansion 
of the Lagrangian, see Eq.~(\ref{action}), to higher order in $\what\kappa$
would be required to consistently assess the impact of the $\Lambda$
term at higher orders in $\what\kappa$. All we can say
at the moment is that, since $\sqrt{-\what g}$ and 
$\sqrt{-\what g_4}$ both involve the field $h_\mu^\mu$ in the same way,
terms involving only $h_\mu^\mu$ will cancel for $V_0-\Lambda=0$.
However, terms that mix $h_\mu^\mu$ and $\phi$, as well as pure $\phi$
terms will survive.  These deserve further analysis.}

A final point is that if $\Lambda\neq 0$ then
one should in principle account for the fact that the background metric in
the bulk in
the absence of matter is not exactly flat, implying that we should
not expand about $\eta_{\what\mu\what\nu}$
except on the brane where we have fine tuned $\Lambda=V_0$.
Indeed, one could ask if it is consistent to presume the existence
of a brane at $y=0$ in the presence of a non-trivial background metric.
However, it can be shown that for $\Lambda<0$ (as required for $\dbar<0$)
the metric is conformally equivalent
to a flat metric for a slice of constant $y$, which we have 
chosen to be at $y=0$.\footnote{We are grateful to S. Carlip for
pointing this out.} In addition, there could be non-trivial dynamics
in the bulk (see for example \cite{chamblin}) that could stabilize
the brane in the required manner.

To summarize, a bulk cosmological
constant does not affect the $\vec n\neq 0$ KK
mode mixing and minimization process but does result in
the vacuum energy seen by gravity being given by $V_0-\Lambda$.
For the natural magnitude of $|V_0|\sim M_S^4$ (as certainly
required for $\dbar<0$), the cancellation between $V_0$ and $\Lambda$
must be essentially exact.  However, we do not regard this as being
unreasonable given that all these quantities will be determined
by the ultimate string theory which might well have such an exact 
cancellation built in by means of symmetry or dynamics.

\section{Implications of \boldmath $V_0\neq0$ for KK Phenomenology}

We have argued that $V_0\neq 0$ is possible even if $\dbar>0$,
and obviously it is required if $\dbar<0$. If $V_0\neq0$,
then the KK phenomenology given in the literature will be altered.  
The crucial point is 
that if $V_0\neq 0$ then the Higgs scalar fields
and the fermionic fields must be rescaled to achieve canonical
normalization. In addition, in the $V_0={1\over 2\dbar}$ minimum
with $\partial V/\partial\Phi\neq0$ there will be Higgs-KK mixing.
Both effects will modify the KK couplings to the physical states.

We first consider whether the KK-mode induced mixing of $s_{\vec n}^1$
with $s$, present at tree-level if $\dbar<0$, could create 
experimentally significant modifications.
We argue that this is not the case since the mixing 
angles $\theta_{\vec n}\sim -{\epsilon\over m_n^2}$ are very small
by virtue of $\eps\propto \kappa$ [see Eq.~(\ref{vtot1})]. It is
only if one performs experiments at energies of order the cutoff
scale $M_S$ that the cumulative effects of these small mixings might
become significant.

As regards rescaling, we have already noted that the vector
fields are not rescaled. However, this is not
the case for fermionic and Higgs fields. In the fermion case,
the Feynman rules read off from the $\call_F^{\vec n}$
of Eq.~(44) of \cite{hanetal} will be modified by a factor of
$(1-{3\over 2}\dbar V_0)^{-1}$. For the $V_0={1\over 2\dbar}$ minimum,
the coupling strengths of the KK modes to $\what\psi
\overline{\what\psi}$ are obtained by multiplying
the Feynman rules of \cite{hanetal} by a factor of 4.
In the Higgs field case, the Feynman rules of \cite{hanetal}
for KK mode coupling to two Higgs fields must be multiplied
by $(1-\dbar V_0)^{-1}$, which is a factor of 2 for the $V_0={1\over 2\dbar}$
minimum. A sampling of the consequences are the following:
\bit
\item
The effective contact interaction of Eq.~(\ref{calaform}) is multiplied
by a factor of 16 for 4-fermion interactions and by a factor
of 4 in the case of vector-vector-fermion-fermion interactions.
This means that the experimental constraints on $M_S$ will
be increased by a factor of 2 ($\sqrt 2$) in the respective cases.
\item
The amplitude for radiating a KK excitation from a fermion (vector boson)
is increased by a factor of 4 (2). This means that the upper bound on
$M_S$  extracted from experimental limits  
must be re-evaluated.  For example, in $\epem\to \gam$+KK mode,
the two contributing diagrams in which the KK excitation is
radiated from the fermion must be multiplied by a factor of 4,
the diagram involving $\epem\to \gam^*\to \gamma$+KK is unchanged
while the $\epem\gam$KK contact term is multiplied by a factor 4.

\eit 

\section{Discussion and Conclusions}

To summarize, we have found that mixing between the Higgs
sector and the KK modes could provide a 
source for electroweak symmetry breaking
even in the absence of tree-level Higgs self interactions. The proposed
mechanism arises automatically if the KK mode sum 
$\sum_{\vec n}{1\over m_{\vec n}^2}\propto \dbar$ is cutoff
at the string scale in such a way that $\dbar<0$. We note that 
electroweak symmetry breaking occurs when $\dbar<0$ whatever
the actual string cutoff scale, $M_S\sim |\dbar|^{-1/4}$,
so long as $M_S$ is sufficiently below $\mplanck$ that the
effective theory we employ can be defined.
Even if the KK modes are not responsible for electroweak symmetry breaking,
the phenomenology of the contact interactions and missing
energy processes which they mediate 
could be substantially modified if the Higgs potential
vacuum expectation value is of order $M_S^4$, as is entirely possible
and perhaps even natural in the string theory context. We have noted
that such a vacuum expectation value does not necessarily lead
to an unacceptably large cosmological constant; it can be canceled
by introducing a cosmological constant in the bulk. As usual,
the cancellation must be fine-tuned.  However, such 
precise cancellation
could be an automatic result of the string theory dynamics or symmetries. 

Undoubtedly, the $\dbar<0$ case is somewhat unusual.  The two major issues
that remain and that we cannot at the moment resolve are the following.
First, we have not explicitly constructed a string theory
for which $\dbar<0$ is predicted. We can only argue that it appears to be
an entirely possible result. Even though the KK states below $M_S$
give a positive contribution to $\dbar$, the way in which
KK modes above $M_S$ couple to the hidden string physics and the 
divergence of the $\sum_{\vec n}{1\over m_{\vec n}^2}$
summation is regulated is highly uncertain. At this time,
one can only say that the result is an effective operator for which
one could have $\dbar<0$.  
Second, since the gravitational corrections are, in the end,
 of order 1 it is possible that results obtained without 
expanding to all orders in $\kappa$ are misleading.
We have not attempted to go beyond the linear expansion, but it is
clear that this issue deserves further investigation. 
We do wish to note that $V_0\neq 0$ does not force us to violate the idea
that we are dealing with an effective theory below the scale $M_S$.
In particular, the Higgs mass and the value of $\Phi$
at the minimum can easily be below $M_S$. Indeed, as shown earlier,
$\what\Phi=\what v=246\gev$ is required for the correct value of $m_W$,
and parameters can be chosen so that
the (tree-level) Higgs mass is also of this same order. For instance,
in the $\dbar<0$ scenario the Higgs mass is of order $\what v$
so long as $m$ of Eq.~(\ref{vform}) is of order $M_S$.  The only requirement
for this to be possible is that  $\Xi$ 
be negative with absolute magnitude of order $M_S^4$. This, we argue,
would not be unnatural in a string theory context.

As regards the $\dbar>0$ case, aside from the $V_0\neq0$
issues already noted above, an interesting new point is the
potential instability at large $\Phi$ implied by Eq.~(\ref{leff3}).
If the $\lambda$ coefficient of the quartic interaction is significantly
larger than 1, $\vtotbar$ becomes negative for $\Phi$ values
significantly below $M_S$ (i.e. such that the effective theory is
likely to still be valid). We have argued that if one enters
the normal $\Phi=v$ minimum early in the evolution of the universe,
then the barrier to penetrating to the large $\Phi$ instability
will be very substantial (of order $M_S^4$). However, a more
detailed investigation is probably called for.

Note added: Ideas concerning possible relations between electroweak
symmetry breaking and gravity have also been considered in the
past\cite{ideas}.

\subsection*{Acknowledgements}
We thank S. Carlip, Z. Lalak, K. Meissner, J. Lykken,
M. Olechowski, J. Pawelczyk, M. Schmaltz and J. Wells for helpful
conversations.
This work was supported in part by the U.S. Department of Energy,
the U.C. Davis Institute for High Energy Physics, the State Committee for
Scientific Research (Poland) under grant No. 2~P03B~014~14 and by Maria
Sklodowska-Curie Joint Fund II 
(Poland-USA) under grant No. MEN/NSF-96-252. 
One of the authors (BG) is indebted
to the U.C. Davis Institute for High Energy Physics for the great
hospitality extended to him while this work was being performed.

\noindent

\end{document}